\documentclass[pra, onecolumn]{revtex4}
\usepackage{graphicx}

\usepackage{ulem}
\usepackage[section]{placeins}
\usepackage{amsmath}
\usepackage{amssymb}
\usepackage{dsfont}
\usepackage{natbib}

\usepackage{subfiles}
\usepackage{graphicx}
\usepackage{array}
\usepackage{calrsfs}
\usepackage{hhline}
\usepackage{relsize}

\begin{document}

\title{A rapid nanometre-precision autocorrelator}

\author{Imogen Morland, Feng Zhu, Paul Dalgarno and Jonathan~Leach$^{*}$}
\address{Institute of Photonics and Quantum Sciences, Heriot-Watt University, David Brewster Building, Edinburgh EH14 4AS, UK}
\address{$^*$j.leach@hw.ac.uk}


\begin{abstract}

The precise measurement of a target depth has applications in biophysics and nanophysics, and non-linear optical methods are sensitive to intensity changes on very small length scales. By exploiting the high sensitivity of an autocorrelator's dependency on path length, we propose a technique that achieves $\approx$30 nm depth resolution for each pixel in 30 seconds. Our method images up-converted pulses from a non-linear crystal using a sCMOS (scientific Complementary Metal–Oxide–Semiconductor) camera and converts the intensity recorded by each pixel to a delay. By utilising statistical estimation theory and using the data from a set of 32$\times$32 pixels, the standard error (SE) of the detected delay falls below 1 nm after 30 seconds of measurement. Numerical simulations show that this result is extremely close to what can be achieved with a shot-noise-limited source and is consistent with the precision that can be achieved with a sCMOS camera.
   
\end{abstract}
\maketitle

\section{\label{sec:Intro}Introduction}

Autocorrelators have applications in measuring the width of ultra-short laser pulses \cite{myslinski1987rapid, oksanen1993femtosecond} and when combined with a spectrometer, the frequency can also be retrieved -- a process known as frequency-resolved optical gating (FROG) \cite{kane1993characterization}. Such systems rely on parametric up-conversion, which exploits the non-linearity of an optical crystal to produce one high energy beam from two lower energy beams \cite{PhysRev.127.1918}.  The process is  subject to the conservation of energy and momentum and has applications in long wavelength imaging systems \cite{midwinter1968image, junaid2019video, yang2020frequency}. Autocorrelators have long held the potential to precisely measure the depth of a target due to the sensitivity of intensity changes to temporal delays between the two beams.

Other sensitive optical methods for measuring depth include Hong-Ou-Mandel (HOM) interferometry \cite{hong1987measurement}, optical coherence tomography (OCT) \cite{huang1991optical} and quantum optical coherence tomography (QOCT) \cite{nasr2003demonstration}. In HOM interferometry, two photons always exit the same arm of a beam splitter when the temporal delay is zero, leading to a dip in coincidence counts. This dip is sensitive to time delays and has been used to measure the group velocity delay of a photon pair to 0.1 fs uncertainty \cite{dauler1999tests,branning2000simultaneous}. This technique was combined with statistical estimation theory to achieve nanometer precision \cite{lyons2018attosecond}. The reported method \cite{lyons2018attosecond} used a dual-arm geometry and achieved the resolution after acquisition times ranging between 1.4 and 15.6 hours.  A similar approach has been used to track the polarisation state of light to $0.01 ^\circ$ precision \cite{harnchaiwat2020tracking}. More recently, HOM interferometry has been used to retrieve images with micrometre-scale depth features \cite{ndagano2021quantum}.

OCT has been used to produce micrometre resolution cross-sectional images of biological systems \cite{huang1991optical}, and has been applied to imaging the eye \cite{fercher1988eye,izatt1994micrometer} as well as other biological materials \cite{clivaz1992high}. Furthermore, QOCT uses HOM interferometry to measure the depth profile from reflective materials and has also resolved biological samples to micrometre resolution \cite{abouraddy2002quantum, nasr2004dispersion, lopez2012quantum}. Whereas, HOM interferometry operates at the single-photon level, leading to low signal intensities and long data acquisition times, autocorrelators can operate at higher intensities and significantly shorter acquisition times. To date, autocorrelators have been used to reach the ultimate quantum limit for measuring incoherent mixtures of ultrashort pulses at the single-photon level \cite{ansari2021achieving}.

\begin{figure}
\centering
\centerline{\includegraphics[]{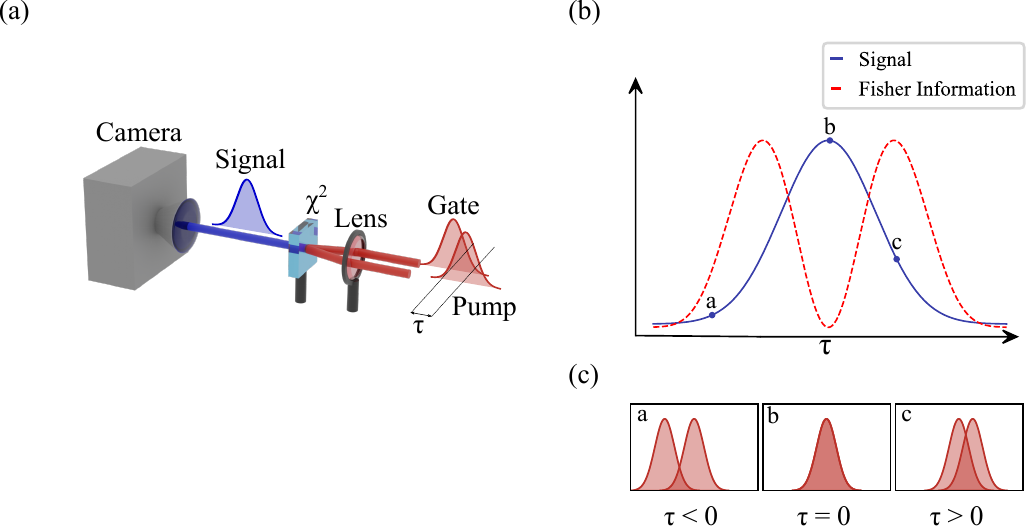}}
\caption{(a) The gate and pump pulses are focused onto a non-linear crystal. The intensity of the up-converted signal is dependent on the temporal delay ($\tau$) between the gate and pump. (b) The signal intensity and Fisher information for different delays between the gate and pump. (c) Gate and pump pulses for three values of $\tau$, the labels corresponds to the data point on plot (b).
\label{Fig: 3D}}
\end{figure}

We build on the prior work and demonstrate a procedure
for measuring the separation of two ultra-short pulses of light to
nanometer resolution within around 30 seconds.  The method uses a sCMOS camera to image the second-harmonic light generated from two pulses with 140 fs pulse width and can therefore be used to record an image corresponding to pulse separation.  We achieve a depth image of dimension 32$\times$32, with each pixel having precision of $\approx$30 nm within 30 seconds of data acquisition.  The data from all pixels can then be combined to provide a single, precise depth estimate, the standard error of which is less than 1 nm after 30 seconds.  The nanometer precision corresponds to a depth accuracy that is more than four orders of magnitude better than the pulse duration of the light (the pulse duration of 140 fs corresponds to a pulse length of 42 $\mu$m).

We follow a four-step procedure in this work: calibration, image acquisition, analysis, and comparison with numerical results.  The first step, calibration, finds the relationship between the changes in the intensity of the generated signal to the delay of the pump pulse with respect the gate pulse for each of the pixels. Here we find the delay between the pulses where changes in the intensity of the signal corresponds to the maximum information content about the pulse separation.  This is found by evaluating the Fisher information for the signal light as a function of the delay.  Each pixel has its own set of calibration parameters.  The data is collected in the second step.  The pulse separation is set to a location close to the maximum Fisher information found in the first step, and we then capture several thousand images of the generated signal on a sCMOS camera. Each 32x32 image contains 1024 pixels with intensity values that can be converted to pulse separations using the information found in the calibration step.  This image acquisition process can be repeated to account for systematic errors in the calibration parameters in the first stage. The third step is the analysis. The standard error (SE) of the pulse separations is calculated for every pixel in the image, and to further improve the resolution, the pulse separation data for all pixels are combined. In the final step, the results are compared to the theoretical limits achievable with a shot-noise limited source.

\section{\label{sec:Theory}Theory}

Our system relies on sum-frequency generation (SFG). This is shown in Fig.~\ref{Fig: 3D} (a), where a pump and gate pulse are focused onto a non-linear crystal and up-converted light is produced. The signal intensity produced depends on the temporal delay $\tau$ between the gate and pump pulses, as shown in Fig.~\ref{Fig: 3D} (b). The Fisher information, which tells us the information
content in the signal changes, is also dependent on $\tau$. The gate and pump pulses required to produce the three signal intensities indicated by dots in Fig.~\ref{Fig: 3D} (b) are shown in Fig.~\ref{Fig: 3D} (c). The right-hand image in Fig.~\ref{Fig: 3D} (c) shows the delay between the gate and pump pulses when the Fisher information is maximum. The signal intensity of light produced by SFG is given by
\begin{equation}
\begin{aligned}
I(\tau) = \int_{-\infty}^{\infty} I_p(t)I_p(t-\tau) \,dt\ ,
\end{aligned}
\label{eqn: I_t}
\end{equation}
\noindent 
where $I_p(t)$ is the intensity of the pump pulse and $I_p(t-\tau)$ is the intensity of the gate pulse. In the case of gate and pump beams with Gaussian temporal profiles, the intensity distribution of the signal pulse is also a Gaussian and is given by
\begin{equation}
\begin{aligned}
I(z) = A \exp{\left(\frac{-z^2}{2 \sigma ^2}\right)} + b ,
\end{aligned}
\label{eqn: I_z}
\end{equation}
\noindent 
where $A$ is the maximum intensity, $z = c\tau$ is the spatial delay between the pulses, $\sigma$ is the standard deviation of the generated signal pulse and $b$ accounts for any background. 

As the intensity of the signal changes as a function of the delay between the gate and the pump, we can use changes in the signal intensity to precisely measure the delay.  The information that is contained in the signal $I(z)$ is given by the Fisher information, which is defined as
\begin{equation}
\begin{aligned}
F(z) = \left(\frac{\partial I(z)}{\partial z}\right)^2 \frac{1}{I(z)}.
\end{aligned}
\label{eqn: Fisher}
\end{equation}
Here we see that more information about the delay is obtained when there is a high relative change in $I(z)$. By substituting Equ. \ref{eqn: I_z} into Equ. \ref{eqn: Fisher}, we derive the Fisher information as a function of the delay to be 
\begin{equation}
\begin{aligned}
F(z) = \frac{A^2 z^2 \exp{\big(\frac{-z^2}{ \sigma ^2}}\big)}{\sigma ^4  \left(A \exp{\big(\frac{-z^2}{2 \sigma ^2}\big)} + b \right) }.
\end{aligned}
\label{eqn: Fisher eval}
\end{equation}
This function is shown in Fig.~\ref{Fig: 3D}(b), and there are two spatial delays where the Fisher information is maximised.  At these locations, changes in the intensity signal carry the most information about the delay between the two pulses.  By deliberately setting the delay to be close or equal to one of these positions we gain the most information from each measurement and maximise the efficiency of our system. 

In our experiment, we use a sCMOS (scientific Complementary Metal–Oxide–Semiconductor) sensor to measure an image of the generated light.  This provides an image where each $i$ and $j$ pixel record an intensity value $I_{i,j}$. After the calibration step, the intensity at each pixel location can be converted to an estimated delay $\tilde{z}_{i,j}$ through the following relationship,
\begin{equation}
\begin{aligned}
\tilde{z}_{i,j} = \sqrt{-2 \sigma_{i,j}^2 \ln{\left(\frac{I_{i,j}-b_{i,j}}{A_{i,j}}\right)}},
\end{aligned}
\label{eqn: z}
\end{equation}
\noindent 
where $b_{i,j}, A_{i,j},$ and $\sigma_{i,j}$ are the corresponding calibration fit parameters for that pixel found in the first stage. Here, we do not consider negative delays since we always select the positive value of $Z_f$.  We take $N$ images with the sCMOS sensor, each image giving an estimate $\tilde{z}_{i,j}$ for each pixel. These estimates are then combined to provide a mean estimated delay for each pixel $\bar{z}_{i, j}$ with an associated standard error SE$_{i, j}$. Data from multiple pixels can be combined to provide a more accurate estimate of the position at the expense of spatial resolution.

Unfortunately, any errors $\Delta \sigma_{i,j}$, $\Delta A_{i,j}$, and $\Delta b_{i,j}$ in the pixel fit parameters lead to unwanted systematic errors in the estimates $\Delta \tilde{z}_{i,j}$.  The combined error in the estimated delay for each pixel can then be calculated using propagation of errors, and is given by
\begin{equation}
\begin{aligned}
\Delta \tilde{z}_{i,j} = \sqrt{ -2 \beta \Delta \sigma ^2 -
\frac{\sigma^2}{2 \beta A^2} \Delta A ^2 - \frac{\sigma^2}{2 \beta (I-b)^2} \Delta b ^2},
\end{aligned}
\label{eqn: error z simp}
\end{equation}
\noindent 
where $\beta = \ln{\Big(\frac{I-b}{A}}\Big)$ is negative. Here we have dropped the $i, j$ subscripts in the formula, but these should be considered. This results in systematic errors in the estimated delays for each pixel, e.g.~a $1\%$ error in the estimates of the parameters leads to an error of order $\approx 600$ nm for the delay estimate for that pixel. As these errors are systematic, they can, however, be accounted for with additional calibration steps.  The errors $\Delta \sigma_{i,j}$, $\Delta A_{i,j}$, and $\Delta b_{i,j}$ can be found from the fit routine in the initial calibration stage.

\section{\label{sec:Set up}Experimental Set-up and Calibration}

\begin{figure}
\centering
\centerline{\includegraphics[]{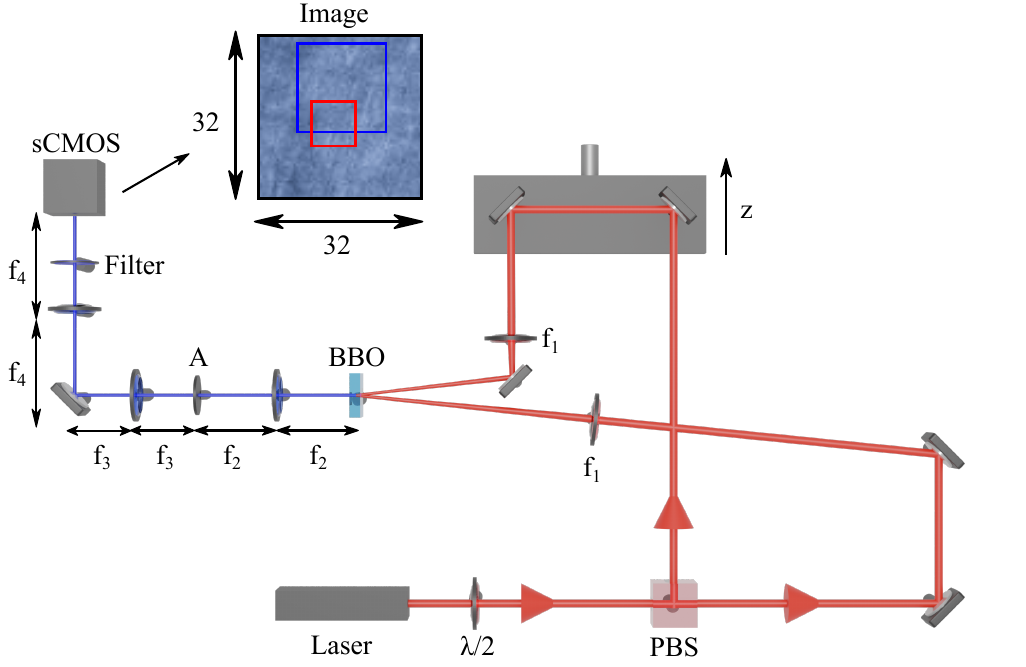}}
\caption{Femtosecond pulses from an 810 nm Ti:sapphire laser are incident on a PBS, and the pump and gate pulses are focused onto a 8x8x1 mm type-2 BBO crystal. The intensity of the pump and gate pulses are controlled via a half-wave plate ($\lambda / 2$) to maximise the upconversion efficiency. The temporal delay between the pulses is varied by scanning a translation stage, and the signal light is imaged by a sCMOS camera (Hamamatsu C11440 ORCA-Flash4.0) via a series of lenses with focal lengths $f_1 = 200$ mm, $f_2 = 30$ mm, $f_3 = 60$ mm and $f_4 = 250$ mm. A 10 nm bandpass filter centered at 405 nm and an aperture blocks any unwanted light before the camera. A 32 x 32 pixel example image of the beam is shown, where the red and blue boxes indicate 8 x 8 and 16 x 16 pixel windows.
\label{Fig: Exp}}
\end{figure}

To achieve our goal of rapid nanometer precision measurements, we built an autocorrelator based on a femtosecond 810 nm Ti:Sapphire laser.  The set-up is shown in Fig.~\ref{Fig: Exp}. Pulses from the laser are incident on a PBS, and the intensity of each output beam is controlled by a half-wave plate ($\lambda / 2$). The pump and gate pulses are focused onto a 8x8x1 mm type-2 Beta Barium Borate (BBO) crystal by two lenses with focal lengths $f_1 = 200$ mm and the signal intensity is varied by scanning a translation stage. Finally, the signal beam is imaged by a sCMOS camera via a series of lenses with focal lengths $f_2 = 30$ mm, $f_3 = 60$ mm and $f_4 = 250$ mm. A 32 x 32 pixel example image of the beam is also shown, where the red and blue boxes indicate 8 x 8 and 16 x 16 pixel windows used in the data acquisition stage.

\begin{figure}[t!]
\centering
\includegraphics[]{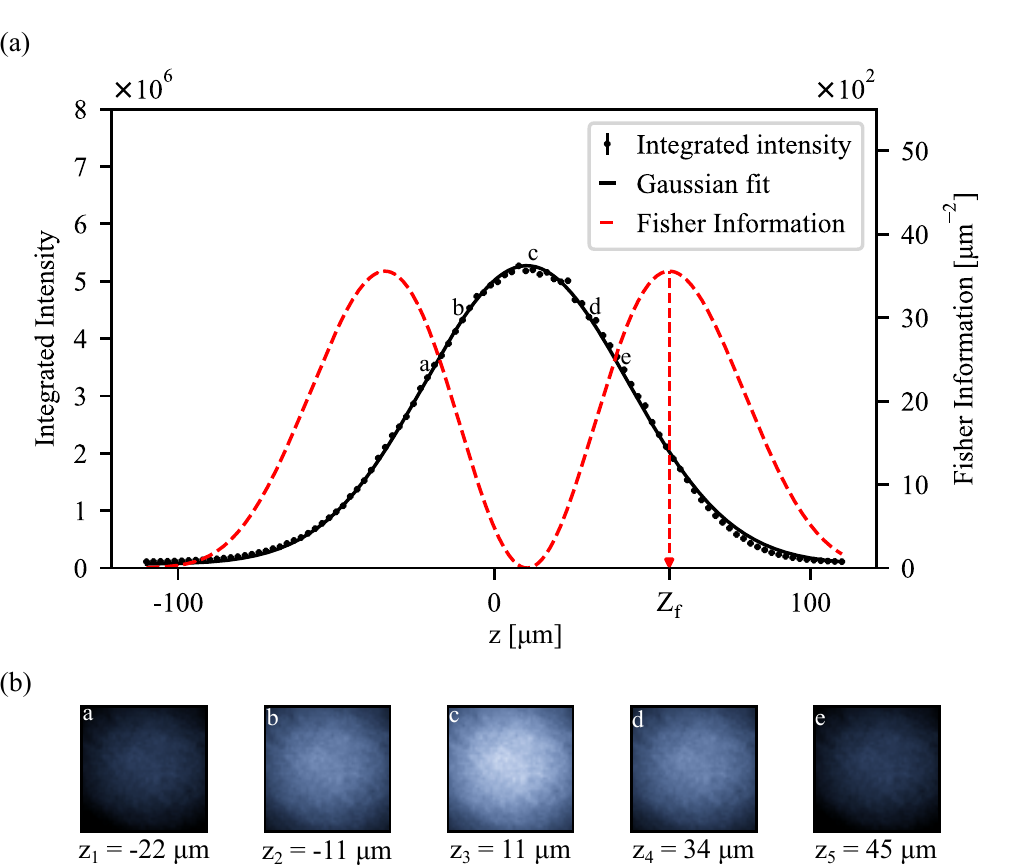}
\caption{(a) Calibration for the integrated intensity, calculated by summing the intensity of every pixel at each stage position, and the associated Fisher information. The positive value of the maximum Fisher information is indicated by $Z_f$. (b) Images of the up-converted beam for five stage positions, indicated by labels in (a).}
\label{fig:calibration}
\end{figure}

Before acquiring data, we must first calibrate the system and find the optimal delay that maximises the Fisher information.  To achieve this, we first record the intensity of a 32 x 32 pixel section of the beam as a function of a translation stage position $I_{xyz}$ for $N_z$ stage positions. We integrate the intensity for each image at each stage position as shown in Fig. \ref{fig:calibration} (a). Then Gaussian fit parameters are extracted to calculate the Fisher information, see Equ.~\ref{eqn: Fisher eval}. The intensity of the signal beam at five stage positions is shown in Fig.~\ref{fig:calibration} (b). The pixel calibration parameters ($A_{i, j}, \sigma_{i, j}, b_{i, j}$) are found at this stage by fitting Gaussians to the data for each pixel.  Once the calibration step is complete, the stage is positioned close to $Z_f$, and we can proceed to the next step.

\begin{figure}
\centering
\centerline{\includegraphics[]{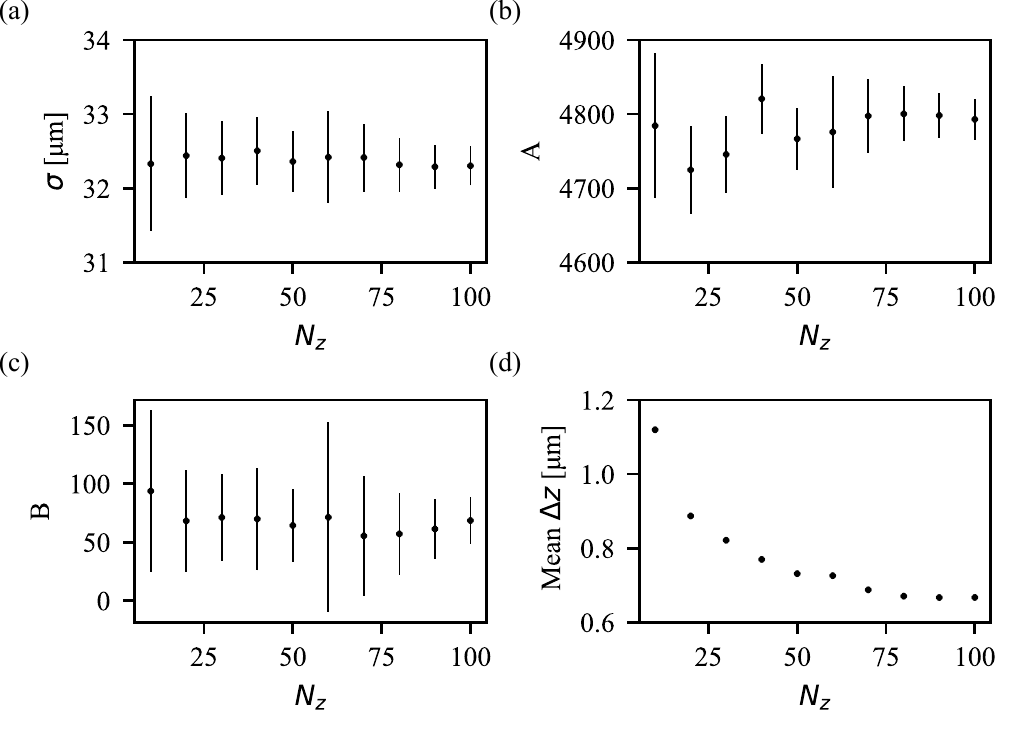}}
\caption{The calibration fit parameters and their associated standard errors for one pixel as a function of the number of positions $N_z$. The standard errors in the calibration fit parameters decreases as $N_z$ increases. (a) The value and standard error in the standard deviation fit parameter for different values of $N_z$. (b) The value and standard error in the amplitude fit parameter for different values of $N_z$. (c) The value and standard error in the background fit parameter for different values of $N_z$. (d) The combined error in the estimated delay, calculated using Equ.~\ref{eqn: error z simp}, decreases as $N_z$ increases. 
\label{Fig: Z_error}}
\end{figure}

As noted above, the error in the pixel parameters introduces a systematic error in the depth estimate.  We can reduce this error by increasing the number of stage positions $N_z$ used for calibration.  We performed the calibration for a range of $N_z$ (between 10 and 100 stage positions in steps of 10) to find the relationship between $N_z$ and the mean error in the estimated delay. A Gaussian function is fitted to each pixel and the standard deviation, maximum amplitude and background fit parameters for one pixel are shown in Fig. \ref{Fig: Z_error} (a)-(c), where the error bars are given by the standard error in the fit parameter. As the number of stage positions increases, the standard error in the fit parameters decreases. The errors for every pixel are then used to calculate the average error in the estimated delay using Equ. \ref{eqn: error z simp}, which follows an exponential decay. This is shown in Fig. \ref{Fig: Z_error} (d), and the error in the estimated delay decreases as $N_z$ decreases. Using this information, we selected 100 stage positions in the final calibration data to minimise the error in the estimated delay.

\section{\label{sec:Sim}Simulation}

\begin{figure}
\centering
\includegraphics[]{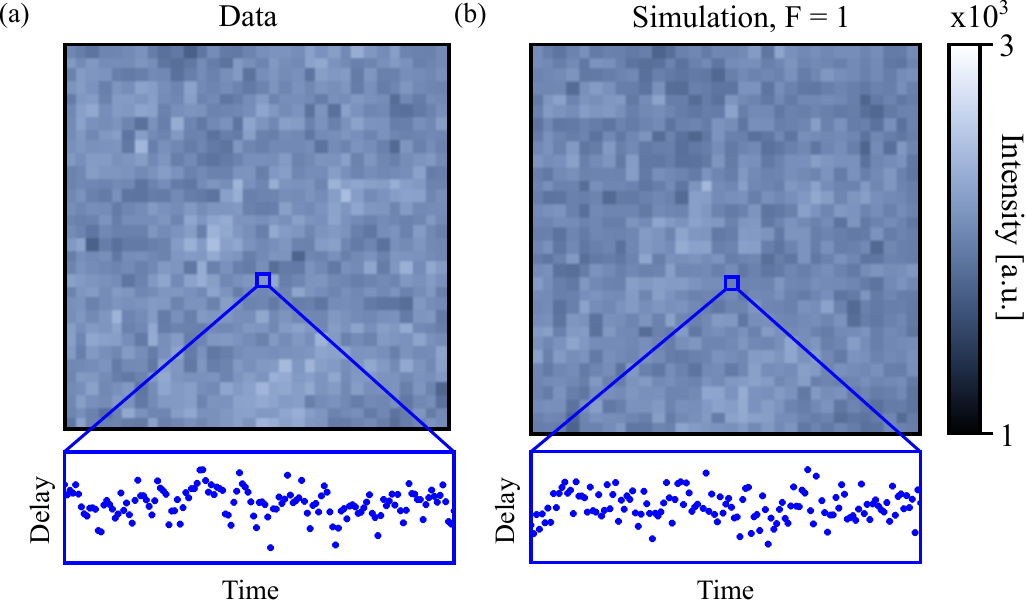}
\caption{Data v Simulation: (a) A 32 x 32 pixel image acquired by the sCMOS camera. The intensity of every pixel is converted to a delay using Equ. \ref{eqn: z} and blue inset shows the brightest pixel's delay as a function of time. (b) A 32 x 32 pixel image taken from the shot-noise limited simulation.}
\label{fig:DataVSim}
\end{figure}

The performance of our system is ultimately limited by the signal intensity fluctuations recorded by every pixel. These are the result of a combination of the shot noise, which cannot be removed, and the sensor noise, which is minimised by optimising the settings.   To compare our experimental results to the theoretical performance that could be achieved with a shot-noise limited detector and a detector performing at the level of the sCMOS camera, we performed several numerical simulations of the experiment. 

We first calculate the average Fano factor \cite{fano1947ionization} across all of the pixels of our recorded data and then generate statistically identical copies of images acquired by the sCMOS camera with the appropriate levels of noise.    The Fano factor evaluates how close the noise in any data is to shot noise and is defined as $F = \sigma_I^2 / \langle I \rangle$, where $\sigma_I^2$ is the variance in the fluctuations and $\langle I \rangle$ is the mean intensity. The factor is unity when the data is shot-noise limited.  For the sCMOS camera used, we measure the average Fano factor across all pixels to be $\bar{F}  = 2.2 \pm 0.1$, and we measure the Fano factor of the brightest pixel to be $F  = 2.3$.

Using the calibration data for each pixel, we calculate the predicted intensity value for every pixel at the location corresponding to the maximum Fisher information. This provides us with an image that we can add noise to. To simulate the intensity fluctuations in the pixels of this image, we then apply a noise function to every $i, j$ pixel in the simulated image according to
\begin{equation}
\begin{aligned}
\tilde{I}_{i,j} =  \sqrt{F_{i,j}}\left(\mathcal{P} ( I_{i,j} )-I_{i,j} \right)+I_{i,j},
\end{aligned}
\label{eqn: I_sim}
\end{equation}
where $\mathcal{P}$ adds Poisson noise to each pixel. This enables us to generate a statistically similar image $\tilde{I}$ to the original image acquired by the sCMOS camera.  The simulated images can have differing levels of noise.  If $F = 1$, they correspond to images that would be shot-noise limited; if $F = 2.2$ they are representative of the images captured by the sCMOS camera in our experiment.  Several thousand statistically similar copies of the sCMOS data can be easily generated by repeatedly applying the noise function to the original fit. These images can then be analysed in the same manner as the data from the real experiment for comparison.

The 32 x 32 pixel intensity profile of one of the images acquired by the sCMOS camera and generated in our shot-noise limited simulation is shown in Fig. \ref{fig:DataVSim} (a) and (b) respectively. The intensity of every $i, j$ pixel is then converted to a delay using Equ. \ref{eqn: z} and the blue window zooms in on the brightest pixel delay. 

\section{\label{sec:Result}Results}

Once the calibration step is finished, we take 1000 frames and calculate the mean estimated delay for each pixel. These mean values are subtracted off the calculated delays in the data acquisition step to remove the systematic error in the estimated delay due to the errors in the calibration fit parameters, see above. In the data acquisition step, we acquire 4400 images of the signal intensity using a sCMOS camera at a location around $Z_f$, each with resolution 32 x 32 pixels. The intensity of every pixel in each image is then converted to an estimated delay using Equ. \ref{eqn: z}.  The standard deviation $\sigma_z$ and standard error $\sigma_z / \sqrt{N}$ in the estimated delay, where $N$ is the number of data points, are calculated. As we continue to acquire more images and $N$ increases the standard error for the estimated delay is updated. We can improve further by utilising the delay estimates for every pixel in each image.  Each image contains 1024 delay estimates, so the standard error of the delay using all the pixels in the image can be reduced by a factor of $\sqrt{1024} = 32$.

\begin{figure}
\centering
\includegraphics[]{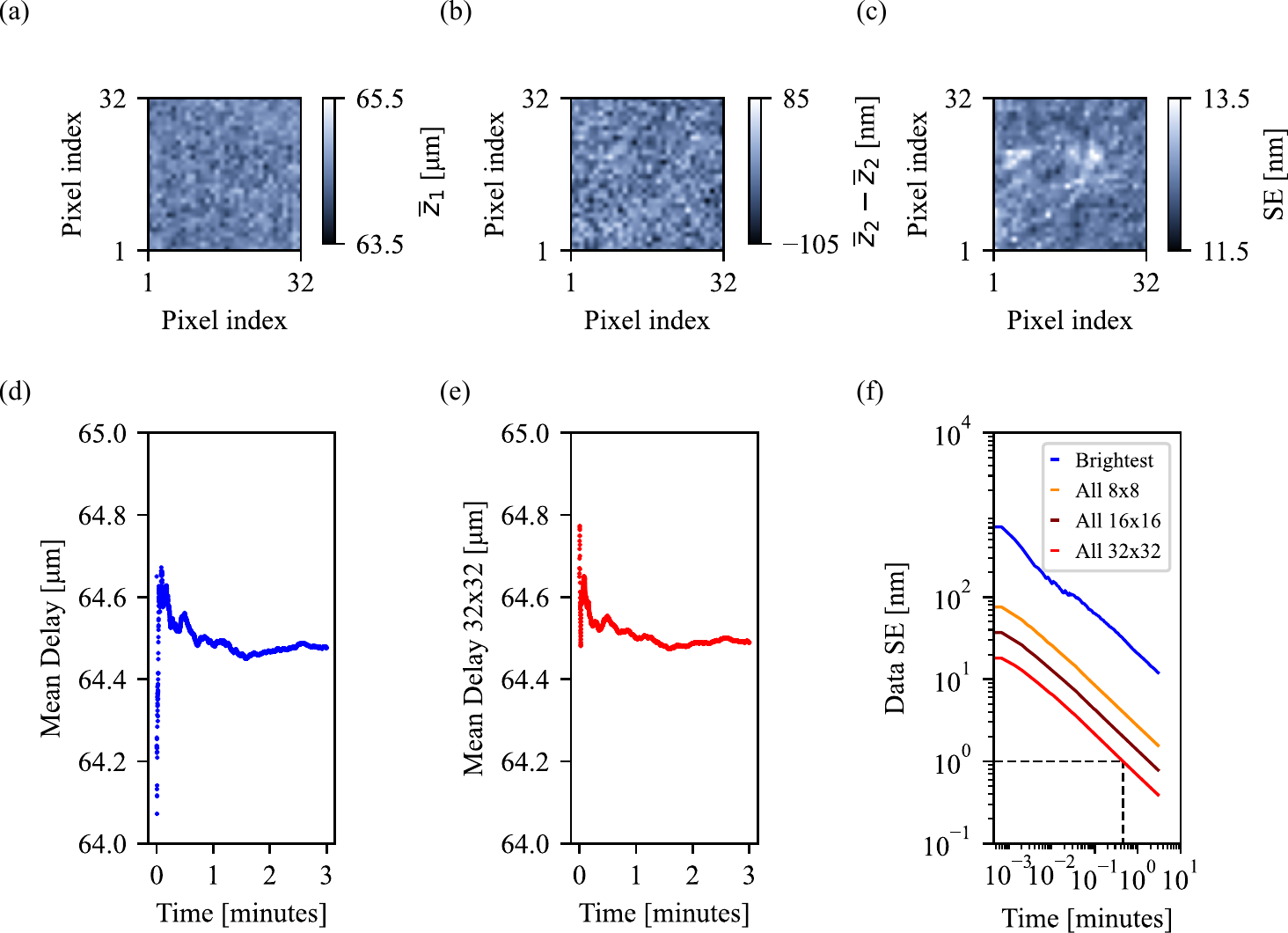}
\caption{(a) The stage is placed at around $Z_f$, and the up-converted intensity is recorded for 1000 frames and converted to an estimated delay $\bar{z}_1$. These estimated delays are subject to a systematic error, and, therefore, the mean estimated delays at the end of the acquisition have a standard deviation of $0.2 ~\mu$m. (b) We repeat the measurement to obtain $\bar{z}_2$ and subtract the values from (a). This corrects for the systematic errors, and the standard deviation of $\bar{z}_2$-$\bar{z}_1$ is now 30 nm. (c) The SE of 32 x 32 pixels at the end of the acquisition. (d) Mean delay for the brightest pixel. (e) Mean delay for 32 x 32 pixels. (f)  SE of the brightest pixel delays and all pixel delays for 3 data windows. The minimum SE for the brightest pixel is 12 nm and the minimum SE for the 8x8, 16x16 and 32 x 32 pixel windows are 1.6 nm, 0.78 nm and 0.39 respectively. The resolution is improved by $\approx$30 times by combining the delay data from 32 x 32 pixels and sub nm resolution is achieved in 27 seconds.}
\label{fig:Results_data}
\end{figure}

The estimated mean pixel delays for the 1000 images are shown in Fig. \ref{fig:Results_data} (a). The standard deviation across all pixels in the image is $0.2 ~\mu$m, with the largest contributing factor to this being the error on the estimated fit parameters. This image can, however, be used as a reference to correct further measurements. A subsequent data set of 4400 frames is then acquired, leading to  corrected mean estimated delays for each pixel shown in Fig. \ref{fig:Results_data} (b). The mean estimated delay after the correction now has a standard deviation across the pixels of 30 nm. Finally, the SE of all pixels at the end of the acquisition is of order 12.5 nm, as shown in Fig. \ref{fig:Results_data} (c).

The mean delay data for the brightest pixel in the image is shown in Fig. \ref{fig:Results_data} (d), where the uncertainty in the delay decreases as the acquisition time increases. Fig. \ref{fig:Results_data} (e) shows the mean delay calculated using all 32 x 32 pixels.  The mean delay here fluctuates significantly less than the individual pixel data. Furthermore, the SE for the brightest pixel and three cropping windows ($8\times8$ pixels, $16\times 16$ pixels, and $32\times32$ pixels) are shown in Fig. \ref{fig:Results_data} (f). The minimum SE achieved for the brightest pixel, 8x8 pixels, 16x16 pixels and 32 x 32 pixels delays are 12 nm, 1.6 nm, 0.78 nm and 0.39 nm respectively. The resolution is improved by $\approx$7 times, $\approx$13 times and $\approx$30 times by combining the delay values for 64 pixels, 256 pixels and 1024 pixels, respectively. Furthermore, we reach sub-nanometer precision in 27 seconds using all 32 x 32 pixels. Combining the delay data of all pixels improves the depth resolution compared to individual pixels, at the cost of losing the transverse resolution.


\begin{figure}
\centering
\includegraphics[]{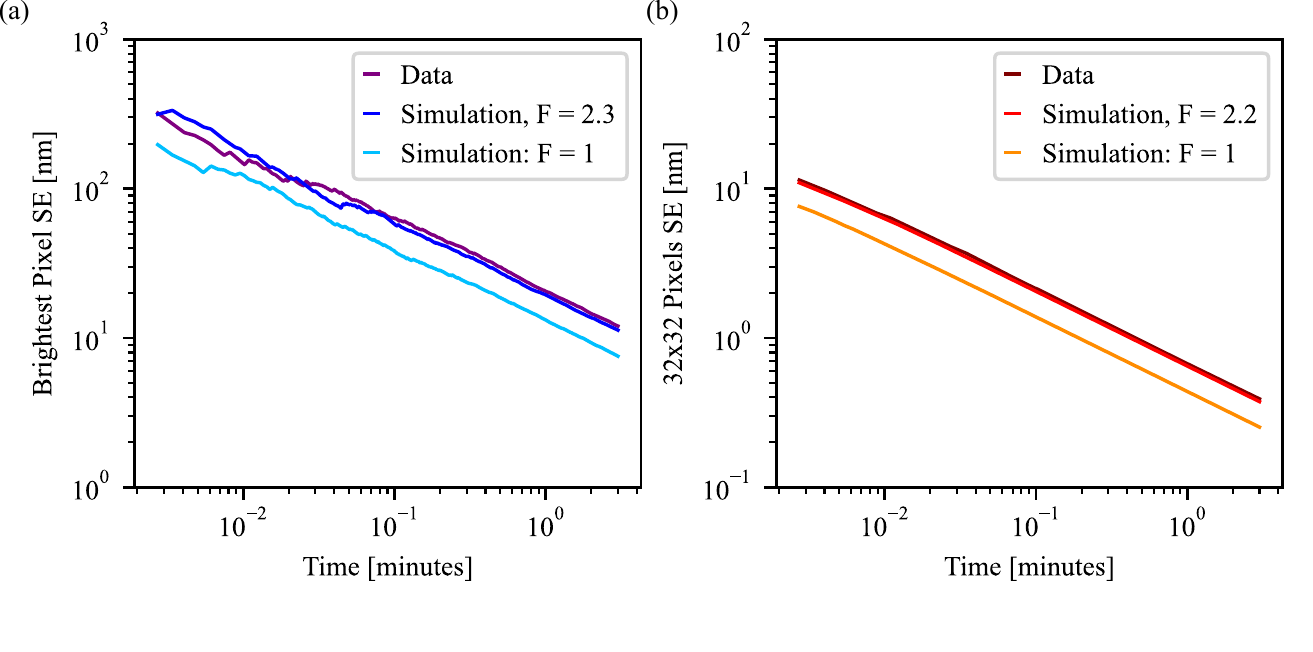}
\caption{(a) SE of the estimated delay for the brightest pixel's data and simulations. The Fano Factor simulation is created by setting the Fano Factor of each pixel to the average Fano Factor of each pixel's sCMOS data. The minimum SE of the brightest pixel’s data is 12 nm, the minimum SE for the F = 2.3 simulation reaches 11 nm, and the minimum SE of the $F = 1$ simulation is 7.5 nm. (b) SE of the delay for 32 x 32 pixels for the data and simulations. The minimum SE in the delay data is 0.39 nm, the simulation with $F = 2.3$ has as a minimum SE of 0.38 nm and the $F = 1$ simulation has a minimum SE of 0.25 nm.}
\label{fig:Data&Sim}
\end{figure}

The experimental data is then compared to the simulation by calculating the standard errors of the brightest pixel and the 32 x 32 pixel crop. Fig. \ref{fig:Data&Sim} (a) shows a comparison between the brightest pixel SE for the data, the shot-noise limited simulation and a simulation with Fano Factor given by the average Fano Factor of the brightest pixel data. In the case of the brightest pixel, $F = 2.3$ and so the simulation is $\sqrt{2.3} = 1.5$ times noisier than the shot-noise limited simulation. The minimum SE of the brightest pixel’s data is 12 nm, the minimum SE for the F = 2.3 simulation reaches 11 nm, and the minimum SE of the shot-noise limited simulation reaches 7.5 nm.  This could be reduced further by increasing the acquisition time. Fig. \ref{fig:Data&Sim} (b) shows the SE of the delay for 32 x 32 pixels for the data, shot-noise limited simulation and a simulation with an average Fano Factor over all pixels of $\bar{F} = 2.2$. The minimum SE of the delay for the data is 0.39 nm, the $\bar{F} = 2.2$ simulation has a minimum SE of 0.38 nm, and the shot-noise limited simulation has a minimum SE of 0.25 nm. Again, the shot-noise limited simulation SE is 1.5 times lower than the simulation with $\bar{F} = 2.2$. On average, the 32 x 32 pixel delay data performs 1.5 times worse than the shot-noise limited simulation and similarly to the $\bar{F}$  simulation. 

The data performs worse than the shot-noise limited simulation because of the additional sensor noise, which we minimised by optimising the sCMOS camera settings. This was done by taking data at different exposure times and upconversion intensities and comparing the Fano Factor and Fourier transform spectrum of the data. We found the Fano Factor was lower for lower intensities and with the exposure time set to 0.04 s. The average Fano factor of the 32 x 32 pixels at the end of the data acquisition was 2.2. Furthermore, the simulation with F = 2.2 performs better than the experimental data due to the drift present in the data increasing the SE. 

\section{\label{sec:Conclusion}Conclusion}

In summary, we propose a method for measuring the separation of optical pulses and achieve $\approx$30 nm depth resolution of each pixel in 30 seconds. Our methods makes use of statistical estimation theory to optimally analyse the up-converted intensity from a non-linear crystal. This is achieved by recording the intensity of up-converted light using a sCMOS camera and converting the intensity recorded by every pixel to a delay estimation. By combining the delay estimations of every pixel in a 32 x 32 image, the SE is calculated to be less than 1 nm after 30 seconds of data acquisition. This precision of measurement has been achieved on significantly faster time scales than previous methods, paving the way for high resolution depth imaging of unknown biological samples.

\section*{Funding}
Science and Technology Facilities Council (ST/S505407/1); Engineering and Physical Sciences Research Council (EP/S001638/1, EP/T00097X/1).

\section*{Disclosures}

The authors declare no conflicts of interest.

\end{document}